\documentclass[11pt,a4paper]{article}
\usepackage{hyperref}
\usepackage[latin1]{inputenc}
\usepackage{dsfont}
\usepackage{graphicx}

\begin{document}
\title{Drop Impact on Liquid Surfaces: \\ Formation of Lens and Spherical Drops at the Air-Liquid Interface}
\author{Ehsan Yakhshi-Tafti, Hyoung J. Cho, Ranganathan Kumar \\
\\\vspace{6pt} Mechanical, Materials and Aerospace Engineering, \\ University of Central Florida, Orlando, FL 32765, USA}
\maketitle
Fluid Dynamics Video: Drop Impact
Droplets at the air-liquid interface of immiscible liquids usually form partially-submerged lens shapes (e.g. water on oil). In addition to this structure, we showed that droplets released from critical heights above the target liquid can sustain the impact and at the end maintain a spherical ball-shape configuration above the surface, despite undergoing large deformation. Spherical drops are unstable and will transform into the lens mode due to slight disturbances.
\\Precision dispensing needles with various tip diameter sizes were used to release pendant drops of deionized water onto the surface of fluorocarbon liquid (FC-43, 3M). A cubic relationship was found between the nozzle tip diameter and the released droplet diameter.  Drop impact was recorded by a high speed camera at a rate of 2000 frames per second. In order for the water drops to sustain the impact and retain a spherical configuration at the surface of the target liquid pool, it is required that they be of a critical size and be released from a certain height; otherwise the commonly observed lens shape droplets will form at the surface.
\\
\\
\\
E. Yakhshi-Tafti, H. J. Cho, and R. Kumar, Journal of Colloid and Interface Science 350, 373-376, 2010
\end{document}